\documentclass[10pt]{article}

\usepackage{scicite}
\usepackage{caption}

\usepackage{times}

\usepackage{graphicx}
\newenvironment{sciabstract}{%
\begin{quote} \bf}
{\end{quote}}

\title{Topological Quantum Computation Based on Chiral Majorana Fermions}

\author
{Biao Lian,$^{1,2\dag}$ Xiao-Qi Sun,$^{2,3\dag}$ Abolhassan Vaezi,$^{2,3}$ \\
Xiao-Liang Qi, $^{2,3,4}$ Shou-Cheng Zhang $^{2,3\ast}$\\
\\
\normalsize{$^{1}$Princeton Center for Theoretical Science, Princeton University,}\\
\normalsize{Princeton, New Jersey 08544-0001, USA}\\
\normalsize{$^{2}$Stanford Center for Topological Quantum Physics, Stanford University,}\\
\normalsize{Stanford, California 94305-4045, USA}\\
\normalsize{$^{3}$Department of Physics, McCullough Building, Stanford University,}\\
\normalsize{Stanford, California 94305-4045, USA}\\
\normalsize{$^{4}$School of Natural Sciences, Institute for Advanced Study,}\\
\normalsize{Princeton, New Jersey 08540, USA}\\
\\
\normalsize{$^\ast$ Correspondence author: S.-C.Z. email: sczhang@stanford.edu}\\
\\
\normalsize{$^\dag$ These authors contribute to the work equally.}
}

\date{}

\begin{document}


\baselineskip24pt


\maketitle


\begin{sciabstract}
Chiral Majorana fermion is a massless self-conjugate fermion which can arise as the edge state of certain two-dimensonal topological matters. It has been theoretically predicted and experimentally observed in a hybrid device of quantum anomalous Hall insulator and a conventional superconductor. Its closely related cousin, Majorana zero mode in the bulk of the corresponding topological matter, is known to be applicable in topological quantum computations. Here we show that the propagation of chiral Majorana fermions lead to the same unitary transformation as that in the braiding of Majorana zero modes, and propose a new platform to perform quantum computation with chiral Majorana fermions. A Corbino ring junction of the hybrid device can utilize quantum coherent chiral Majorana fermions to implement the Hadamard gate and the phase gate, and the junction conductance yields a natural readout for the qubit state.
\end{sciabstract}

% In setting up this template for *Science* papers, we've used both
% the \section* command and the \paragraph* command for topical
% divisions.  Which you use will of course depend on the type of paper
% you're writing.  Review Articles tend to have displayed headings, for
% which \section* is more appropriate; Research Articles, when they have
% formal topical divisions at all, tend to signal them with bold text
% that runs into the paragraph, for which \paragraph* is the right
% choice.  Either way, use the asterisk (*) modifier, as shown, to

Chiral Majorana fermion, also known as Majorana-Weyl fermion, is a massless fermionic particle being its own antiparticle proposed long ago in theoretical physics. The simplest chiral Majorana fermion is predicted in 1 dimensional (1D) space, where it propagates unidirectionally. In condensed matter physics, 1D chiral Majorana fermions can be realized as quasiparticle edge states of a 2D topological state of matter \cite{qi2011}. A celebrated example is the $p+ip$ chiral topological superconductor (TSC), which carries a Bogoliubov-de Gennes (BdG) Chern number $\mathcal{N}=1$, and can be realized from a quantum anomalous Hall insulator (QAHI) with Chern number $\mathcal{C}=1$ in proximity with an $s$-wave superconductor \cite{qi2010b,chung2011,wang2015a,strubi2011}. A QAHI-TSC-QAHI junction implemented this way is predicted to exhibit a half quantized conductance plateau induced by chiral Majorana fermions \cite{chung2011,wang2015a}, which has been recently observed in the Cr doped (Bi,Sb)$_2$Te$_3$ thin film QAHI system in proximity with Nb superconductor \cite{he2017}. Chiral Majorana fermion could also arise in the Moore-Read state of fractional quantum Hall effect \cite{moore1991} and topologically ordered states of spin systems \cite{kitaev2006}.

A closely related concept, Majorana zero modes (MZMs) which emerge in the bulk vortices of a $p+ip$ TSC \cite{read2000} or at the endpoints of a 1D $p$-wave TSC \cite{kitaev2001,lutchyn2010}, are known to obey non-Abelian braiding statistics and can be utilized in fault-tolerant topological quantum computations \cite{ivanov2001,kitaev2003,alicea2011,alicea2012,aasen2016,karzig2017}. Despite the theoretical progress made during the past decade on employing MZMs in universal quantum computation \cite{alicea2011,alicea2012,aasen2016,karzig2017}, due to the localized and point-like nature of MZMs, all existing proposed architectures inevitably require nano-scale design and control of the coupling among MZMs.
%mechanical movements of MZMs and involve nano-scale design and control protocol of quantum gates.
As an essential step towards topological quantum computing, the braiding of MZMs has not yet been experimentally demonstrated.

In this paper, we propose a novel platform to implement topologically protected quantum gates at mesoscopic scales, which utilizes propagation of chiral Majorana fermions with purely electrical manipulations instead of MZMs.

\section*{Chiral Majorana Fermion Qubits}
\begin{figure}[tbp]
\begin{center}
\includegraphics[width=3.5in]{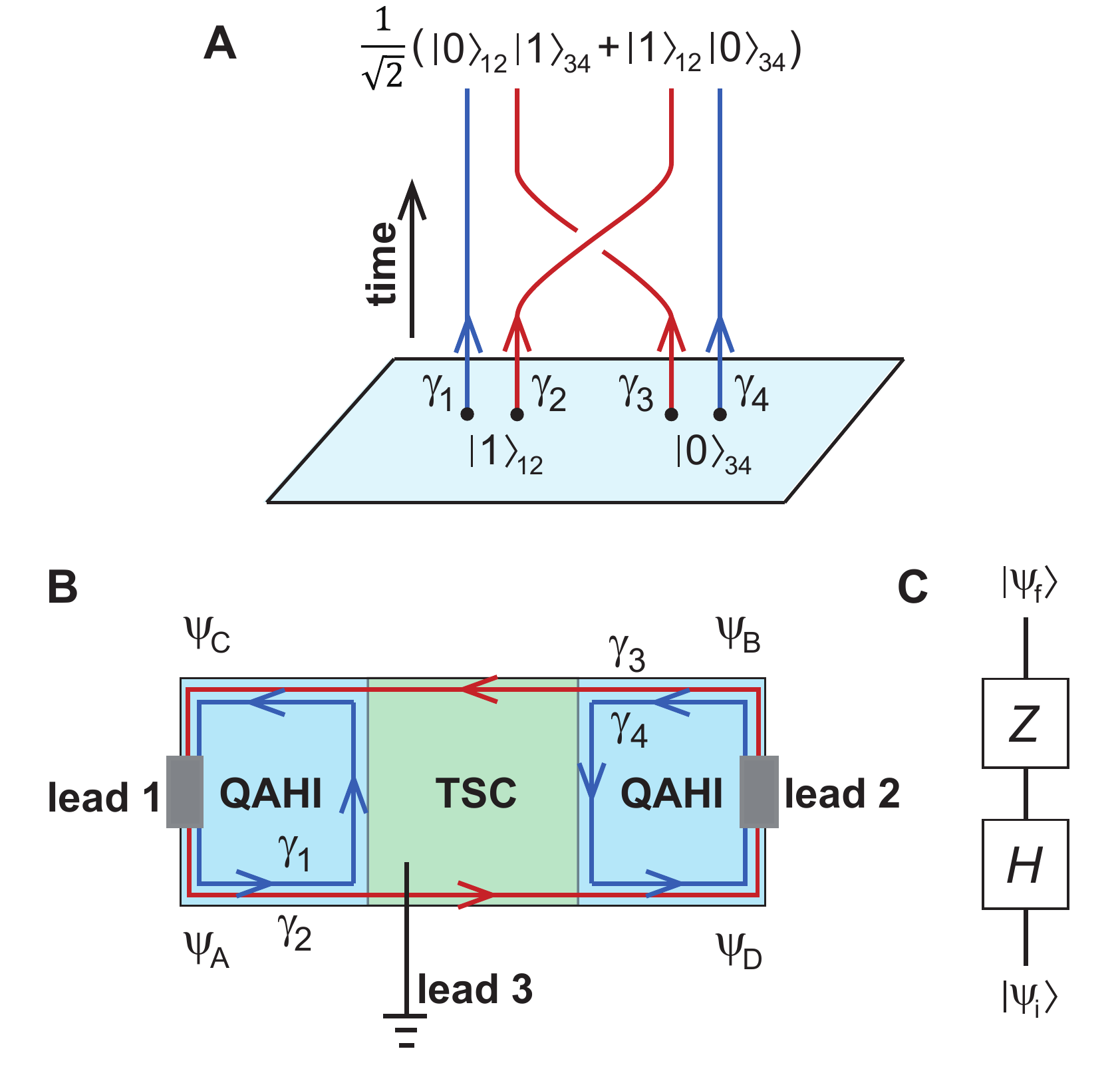}
\end{center}
\caption{(A) The braiding of vortices in $p+ip$ TSC. Each two vortices support two states of a single fermion, and the braiding leads to a non-Abelian operation and maps a product state of vortices $12$ and $34$ into an entangled state, as a consequence of exchanging MZMs $\gamma_2,\gamma_3$. (B) Our proposed device of QAHI-TSC-QAHI junction. The same partner switch as in (A) occurs between incoming electrons from A, B and outgoing electrons in C, D. (C) Such a exchange leads to a non-Abelian gate that is equivalent to a Hadamard gate $H$ followed by a Pauli-Z gate $Z$. %(E) Evolution of entanglement entropy $S_E$ between left and right halves of the junction (divided by dashed line in Fig. \ref{fig1}A) with time $t$ (arbitrary unit) after an electron above the fermi sea is injected from lead $1$, where $S_{E0}$ is the entanglement entropy of the fermi sea.
}
\label{fig1}
\end{figure}

The main goal of our proposal is to show that the chiral Majorana fermion edge state of TSC can be used to realize non-Abelian quantum gate operations on electron states, even if there is no non-Abelian anyon travelling along the edge. Since our proposal is closely related to the braiding of MZMs in vortices of $p+ip$ TSC, we begin by reviewing this process, as is illustrated in Fig. \ref{fig1} A. Each vortex supports a single MZM $\gamma_i$, and thus two vortices together defines two quantum states of a fermion degree of freedom. The MZM operators satisfy the anticommutation relation $\{\gamma_i,\gamma_j\}=\delta_{ij}$. If we define $f_{12}=\frac12\left(\gamma_1+i\gamma_2\right)$ as a complex fermion number, the two states are labeled by $f_{12}^\dagger f_{12}=0,1$, which corresponds to $i\gamma_1\gamma_2=-1,+1$ respectively. When two vortices are exchanged, the corresponding MZMs also got exchanged. In the process in Fig. \ref{fig1} A, we have $\gamma_2\rightarrow \gamma_3,~\gamma_3\rightarrow -\gamma_2$. The relative minus sign is necessary to preserve the fermion number parity $i\gamma_2\gamma_3$ of this pair. As a consequence, the eigenstates of $i\gamma_1\gamma_2$ and $i\gamma_3\gamma_4$ evolves to eigenstates of $i\gamma_1\gamma_3$ and $-i\gamma_2\gamma_4$, which are entangled states when written in the original basis of $i\gamma_1\gamma_2$ and $i\gamma_3\gamma_4$. For example, the state $|1\rangle_{12}|0\rangle_{34}$ evolves into $\frac1{\sqrt{2}}\left(|0\rangle_{12}|1\rangle_{34}+|1\rangle_{12}|0\rangle_{34}\right)$. Since the vortices have long range interaction, the Abelian phase during the braiding may not be well-defined, but the non-Abelian unitary operation is robust\cite{ivanov2001}. From the reasoning presented above, one can see that the non-Abelian gate during MZM braiding is a direct consequence of exchanging MZMs $\gamma_2,\gamma_3$. The resulting gate must be non-Abelian because $i\gamma_1\gamma_2$ anticommutes with $i\gamma_1\gamma_3$. Therefore the same non-Abelian gate can be realized by other physical process that exchanges Majorana fermions, even if no braiding of non-Abelian anyon is involved. In the following, we will show how to obtain a new realization of the same gate by making use of chiral Majorana fermion edge states of TSC and complex chiral fermion edge states of QAHI.

The device we propose to study is a 2D QAHI-TSC-QAHI junction predicted in Refs. \cite{chung2011,wang2015a}. As is shown in Fig. \ref{fig1} B,
%, which is derived from generalized Landauer-B\"uttiker formula in Refs. \cite{chung2011,wang2015a},
the junction consists of two QAHI\cite{liu2008,yu2010,chang2013b} of Chern number $\mathcal{C}=1$ and a chiral TSC of BdG Chern number $\mathcal{N}=1$. The conductance $\sigma_{12}$ is measured between metallic leads $1$ and $2$ by driving a current $I$, where no current flows through lead $3$ which grounds the TSC. Each edge between the chiral TSC and the vacuum or a QAHI hosts a chiral Majorana fermion edge mode governed by a Hamiltonian $H_{M}(x)=-i\hbar v_F\gamma(x)\partial_x\gamma(x)$, where $\gamma(x)$ is the Majorana operator satisfying $\gamma(x)=\gamma^\dag(x)$ and the anti-commutation relation $\{\gamma(x),\gamma(x')\}=\delta(x-x')/2$, $v_F$ is the Fermi velocity, and $x$ is the coordinate of the 1D edge.
In contrast, each edge between a QAHI and the vacuum hosts a charged chiral fermion (electron) edge mode with a Hamiltonian $H_{F}(x)=-i\hbar v_F\psi^\dag(x)\partial_{x}\psi(x)$, where $\psi(x)$ and $\psi^\dag(x)$ are the annihilation and creation operators of the edge fermion, and we have assumed chemical potential $\mu=0$ for the moment. By defining two Majorana operators $\gamma_1=(\psi+\psi^\dag)/2$ and $\gamma_2=(\psi-\psi^\dag)/2i$ (hereafter $\gamma_i=\gamma_i(x)$ is short for chiral Majorana fermion), one can rewrite $H_F(x)$ as $H_F(x)=-i\hbar v_F (\gamma_1\partial_x\gamma_1+\gamma_2\partial_x\gamma_2)$, which implies a charged chiral fermion mode is equivalent to two chiral Majorana fermion modes. As a result, the edge states of the junction consist of four chiral Majorana fermion modes $\gamma_i$ ($1\le i\le4$) as shown in Fig. \ref{fig1}B, which are related to the charged chiral fermion modes on the QAHI edges as $\psi_A=\gamma_1+i\gamma_2$, $\psi_B=\gamma_4+i\gamma_3$, $\psi_C=\gamma_1-i\gamma_3$ and $\psi_D=\gamma_4+i\gamma_2$ \cite{chung2011}.

%Our key observation is the paths of four chiral Majorana modes $\gamma_i$ in Fig. \ref{fig1}A are topologically equivalent to an exchange (braiding) of $\gamma_2$ and $\gamma_3$ as shown in Fig. \ref{fig1}B and Fig. \ref{fig1}C, which takes $\gamma_2\rightarrow -\gamma_3$ and $\gamma_3\rightarrow \gamma_2$ \cite{nayak2008}. This braiding turns the two incident charged chiral fermion modes $\psi_A$ and $\psi_B$ into the two outgoing modes $\psi_C$ and $\psi_D$.
Our key observation is that the same kind of partner switch of Majorana fermions as the vortex braiding occurs in this device between incoming and outgoing electrons. An incoming electron from lead $A$ becomes a nonlocal fermion simultaneously on the two edges of TSC described by $\gamma_1$ and $\gamma_2$. If we measure the number of outgoing electrons in leads $C$ or $D$, we find that the outgoing states in the two leads are entangled, because the number operators in these leads do not commute with those of incoming electrons.

To be more specific and to make a connection with quantum computation, consider the low current limit $I\rightarrow0$ where electrons are injected from lead $1$ one by one, each of which occupies a travelling wave packet state of $\psi_A$. The occupation number $0$ or $1$ of such a fermion wave packet state then defines a qubit $A$ with basis $|0_A\rangle$ and $|1_A\rangle$. Similarly, we can define the qubits $B$, $C$ and $D$ for $\psi_B$, $\psi_C$ and $\psi_D$, respectively. At each moment of time, the real and imaginary parts of the fermionic annihilation operator of each wave packet state define two self-conjugate Majorana operators localized at the wave packet. When the wave packets move out the superconducting region, they merge with a different partner and form states of the outgoing qubits. %As time $t$ increases, these local Majorana operators move along the paths of $\gamma_i$, and yield four braiding world lines as shown in Fig. \ref{fig1}B, thus naturally obey the same braiding statistics as that of MZMs.
In the evolution of the incident electrons, qubits $A$ and $B$ span the Hilbert space of the initial state $|\psi_i\rangle$, while qubits $C$ and $D$ form the Hilbert space of the final state $|\psi_f\rangle$. In the same way as the MZM braiding case, the exchange of $\gamma_2$ with $\gamma_3$ then leads to a unitary evolution
\begin{equation}
\left(\begin{array}{c}|0_C0_D\rangle\\|0_C1_D\rangle\\|1_C0_D\rangle\\|1_C1_D\rangle\end{array}\right) =\frac{1}{\sqrt{2}} \left(\begin{array}{cccc}1&0&0&1\\0&1&1&0\\0&-1&1&0\\-1&0&0&1\end{array}\right) \left(\begin{array}{c}|0_A0_B\rangle\\|0_A1_B\rangle\\|1_A0_B\rangle\\|1_A1_B\rangle\end{array}\right)\ .
\end{equation}
This transformation should be viewed as an S-matrix between incoming and outgoing electron states. Note that the fermion parity is conserved in the unitary evolution. If we define a new qubit $(|0\rangle,|1\rangle)$ in the odd fermion parity subspace as $(|0_A1_B\rangle, |1_A0_B\rangle)$ initially and $(|0_C1_D\rangle, |1_C0_D\rangle)$ at the final time, the above unitary evolution is exactly a topologically protected Hadamard gate $H$ followed by a Pauli-Z gate $Z$ as shown in Fig. \ref{fig1}C, namely, $|\psi_f\rangle=ZH|\psi_i\rangle$, where
\begin{equation}
H=\frac{1}{\sqrt{2}} \left(\begin{array}{cc}1&1\\1&-1\end{array}\right)\ ,\quad Z=\left(\begin{array}{cc}1&0\\0&-1\end{array}\right)\ .
\end{equation}
The same conclusion holds for the even fermion parity subspace. Therefore, the two qubits A and B (C and D) behaves effectively as a single qubit, and we can regard qubit A (C) as the data qubit, while qubit B (D) is a correlated ancilla qubit.

\begin{figure}[tbp]
\begin{center}
\includegraphics[width=3.5in]{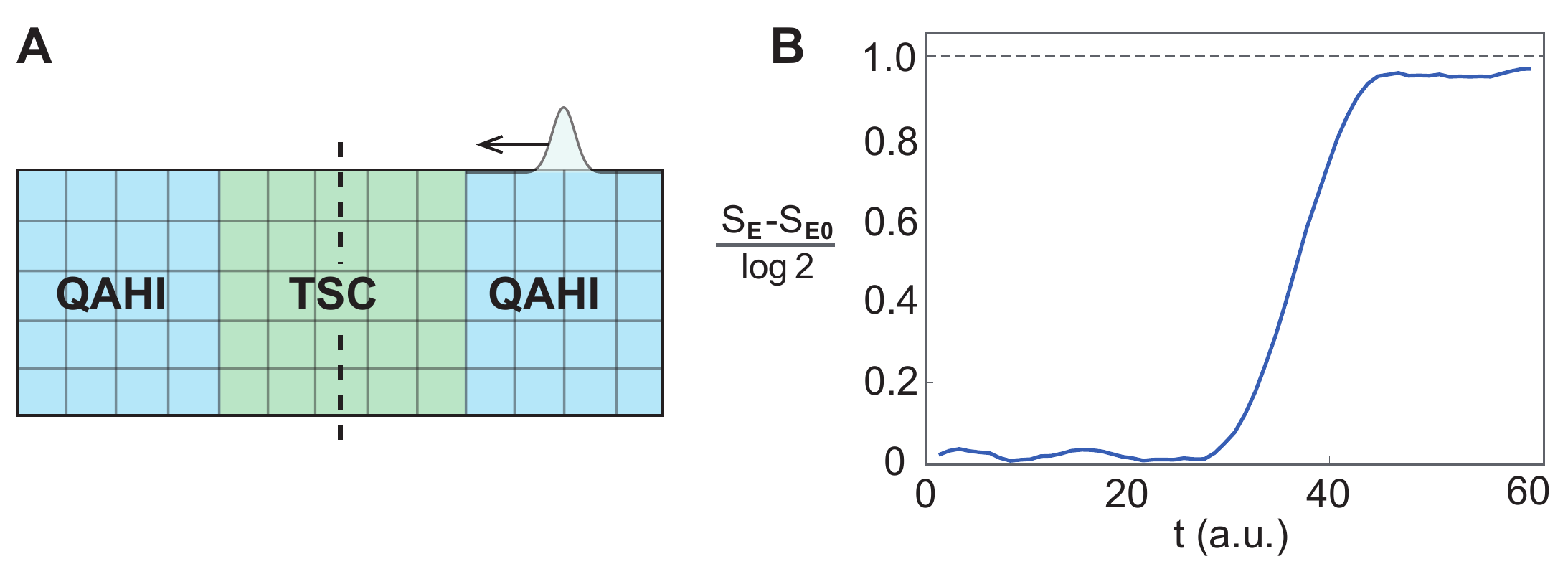}
\end{center}
\caption{(A) The setup for numerical computation of entanglement entropy. We use a lattice model of QAHI-TSC-QAHI junction, add an initial edge wave packet on a QAHI edge, and then examine the time evolution of the state and the entanglement entropy between left and right part of the lattice separated by the dashed line.
%\textcolor{red}{(Biao please add more description if needed.)}
(B) Evolution of entanglement entropy $S_E$ between left and right halves of the junction (divided by dashed line in (A)) with time $t$ (arbitrary unit) after an electron above the fermi sea is injected from lead $1$, where $S_{E0}$ is the entanglement entropy of the fermi sea.}
\label{newfig2}
\end{figure}

For an electron incident from lead $1$ represented by initial state $|\psi_i\rangle=|1_A0_B\rangle$, the junction turns it into a final state $|\psi_f\rangle=(|0_C1_D\rangle+|1_C0_D\rangle)/\sqrt{2}$. This implies \cite{suppl} that the entanglement entropy between left and right halves of the junction divided by the dashed line in Fig. \ref{newfig2}A increases by $\log 2$. Indeed, this is verified by our numerical calculation in a lattice model of the junction (Fig. \ref{newfig2}A), where the entanglement entropy $S_E$ increases with time $t$ as shown in Fig. \ref{newfig2}B, after an electron is injected from lead $1$ above the fermi sea. More details of this calculation is provided in the supplementary material \cite{suppl}. Since $\psi_C$ and $\psi_D$ propagate into leads $1$ and $2$, respectively, the electron has $r=1/2$ probability to return to lead $1$, and $t=1/2$ probability to tunnel into lead $2$. This yields \cite{chung2011} a half-quantized two-terminal conductance $\sigma_{12}=te^2/h=e^2/2h$. Since lead $1$ (lead $2$) connects $\psi_A$ ($\psi_B$) with $\psi_C$ ($\psi_D$) (Fig. \ref{fig1}B), we are in fact identifying the charge basis of final qubit $C$ ($D$) with that of initial qubit $A$ ($B$). Accordingly, the conductance $\sigma_{12}$ provides a natural measurement of the overlap probability between $|\psi_i\rangle$ and $|\psi_f\rangle$ under this common basis, namely, $\sigma_{12}=(1-|\langle\psi_f|\psi_i\rangle|^2)e^2/h$.

As we have discussed, the above process is topologically equivalent to fusion and braiding of four vortex operators in the TSC bulk \cite{suppl,bonderson2006}. More concretely, when the electron of an incident state $|1_A0_B\rangle$ reaches the boundary of the TSC, one can imagine an operation of dragging the electron (fermion) into the Hilbert space of two nearby vortices $\sigma_1$ and $\sigma_2$ in the TSC bulk, after which $\sigma_1$ and $\sigma_2$ are in the fermionic fusion channel. Meanwhile, one can create another two vortices $\sigma_3$ and $\sigma_4$ in the bulk of TSC in the vacuum fusion channel. Next, one can braid the vortices, fuse $\sigma_1$ with $\sigma_3$, and $\sigma_2$ with $\sigma_4$. Lastly, one can drag the state in the Hilbert space of $\sigma_1$ and $\sigma_3$ onto the QAH edge of $\psi_C$, and that of $\sigma_2$ and $\sigma_4$ onto the QAH edge of $\psi_D$. During such a vortex braiding and fusing process, there is no Majorana fermion propagating on the TSC edge. However, the initial state and final state in this case are the same as above process of chiral Majorana fermion propagation \cite{suppl}, so the two processes are topologically equivalent.

\section*{A Testable Quantum Gate}

The conductance $\sigma_{12}$ of the above junction, however, cannot tell whether chiral Majorana fermions $\gamma_i$ are coherent or not during the propagation, and thus whether the process is a coherent quantum gate. For instance, if a random phase factor is introduced in the propagation of $\psi_C$ and $\psi_D$, a pure initial state $|\psi_i\rangle=|1_A0_B\rangle$ will evolve into a mixed final state with a density matrix $\rho_f=(|0_C1_D\rangle\langle0_C1_D|+|1_C0_D\rangle\langle1_C0_D|)/2$, while the conductance remains $\sigma_{12}=[1-\mbox{tr}(\rho_f|\psi_i\rangle\langle\psi_i|)]e^2/h=e^2/2h$.

\begin{figure}[tbp]
\begin{center}
\includegraphics[width=4in]{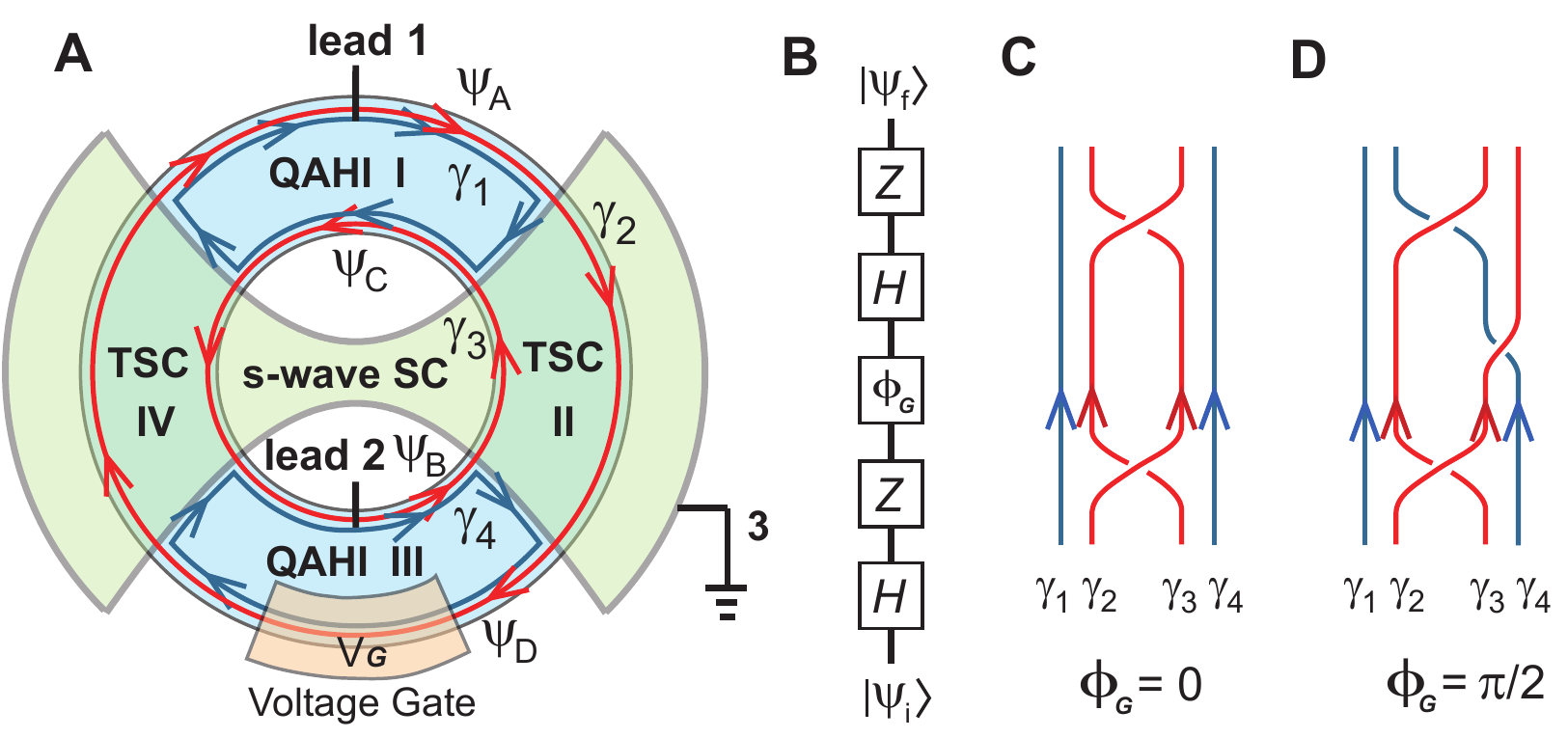}
\end{center}
\caption{Quantum interference in the QAHI-TSC-QAHI-TSC Corbino junction. (A) The Corbino junction consists of a Corbino QAHI ring with a fan-shaped $s$-wave superconductor on top of it which drives regions II and IV into TSC, and a voltage gate $V_G$ is added at the bottom edge. (B) Such a junction is equivalent to a series of single-qubit quantum gates $ZHR_{\phi_G}ZH$, where $R_{\phi_G}$ is a phase gate controlled by $V_G$. (C) and (D) The MZM braiding process that gives the same gates as the corbino device with $\phi_G=0$ and $\phi_G=\pi/2$, respectively.}
\label{fig2}
\end{figure}

To tell whether the system as a quantum gate is coherent, we propose to implement a Corbino geometry QAHI-TSC-QAHI-TSC junction as shown in Fig. \ref{fig2}A, and measure the conductance $\sigma_{12}$ between lead $1$ and lead $2$. The junction can be realized by attaching a fan-shaped $s$-wave superconductor on top of a $\mathcal{C}=1$ QAHI Corbino ring, with a proper out-of-plane magnetic field driving the two regions II and IV into the $\mathcal{N}=1$ TSC phase \cite{wang2015a,he2017}. A voltage gate $V_G$ is added on the bottom edge of QAHI region III covering a length $l_G$ of the edge. Lead $3$ grounds the superconductor and has no current passing through. At zero gate voltage, the edge states of the Corbino junction are four chiral Majorana edge states $\gamma_i$ ($1\le i\le4$) as shown in Fig. \ref{fig2}A.

The gate voltage $V_G$ on the bottom edge of region III behaves as a chemical potential term $H_G=eV_G\psi^\dag_D\psi_D$ for $\psi_D=\gamma_4+i\gamma_2$ in a length $l_G$. In the language of quantum computation, this induces a phase gate
\begin{equation}
R_{\phi_G}=\left(\begin{array}{cc}e^{-i\phi_G}&0\\0&1\end{array}\right)
\end{equation}
acting on the corresponding qubit $D$, where the phase shift $\phi_G=eV_Gl_G/\hbar v_F$ is tunable via $V_G$. Accordingly, the fermion operator $\psi_D$ undergoes a unitary evolution $\psi_D\rightarrow e^{i\phi_G}\psi_D$. In particular, when $\phi_G=\pi/2$, this is equivalent to an exchange of Majorana modes $\gamma_2$ and $\gamma_4$, namely, $\gamma_4\rightarrow\gamma_2$, and $\gamma_2\rightarrow-\gamma_4$.

If we regard the charged chiral edge modes of QAHI region I ($\psi_A$ and $\psi_C$) as the data qubit, and those of QAHI region III ($\psi_B$ and $\psi_D$) as the ancilla qubit, the junction can be viewed as a series of quantum gates as shown in Fig. \ref{fig2}B, with a total unitary evolution $|\psi_f\rangle=ZHR_{\phi_G}ZH|\psi_i\rangle$. Fig. \ref{fig2}C and \ref{fig2}D show the MZM braiding processes that results in the same non-Abelian gate as the $\phi_G=0$ and $\pi/2$ case, respectively. For an electron incident from lead $1$ represented by the initial state $|\psi_i\rangle=|1_A0_B\rangle$, the finial state is
\begin{equation}
|\psi_f\rangle=e^{-i\phi_G/2}\left(\cos\frac{\phi_G}{2}|0_A1_B\rangle+ i\sin\frac{\phi_G}{2}|1_A0_B\rangle\right)\ .
\end{equation}
Therefore, the two-terminal conductance of this Corbino junction is
\begin{equation}
\sigma_{12}=(1-|\langle\psi_f|\psi_i\rangle|^2)\frac{e^2}{h}=\frac{1+\cos\phi_G}{2}\frac{e^2}{h}\ ,
\end{equation}
which oscillates as a function of $V_G$ with a peak-to-valley amplitude $e^2/h$. In contrast, if the system loses coherence completely, the final state will be the maximally mixed state described by density matrix $\rho_f=(|0_A1_B\rangle\langle0_A1_B|+|1_A0_B\rangle\langle1_A0_B|)/2$, and the conductance will constantly be $\sigma_{12}=e^2/2h$. Therefore, the oscillation amplitude of $\sigma_{12}$ measures the coherence of the chiral Majorana fermions in the junction.

So far we have assumed chemical potential $\mu=0$ on all QAHI edges except the interval covered by voltage gate. In general, $\mu$ is nonzero, and is nonuniform along the QAHI edges when there are disorders. Such a nonzero landscape of $\mu$ contributes an additional phase gate, which leads to a phase shift $\phi_G\rightarrow\phi_G+\phi_0$, with $\phi_0$ being a fixed phase \cite{suppl}. Experimentally, the gate voltage $V_G$ and thus $\phi_G$ can be well controlled by current techniques at a high precision level \cite{an2011}.
%Besides, when odd number of magnetic $\pi$ flux vortices are inserted into region II (IV) in Fig. 2A, the sign of $\gamma_2$ will flip sign when propagating along the edge of region II (IV). As a result, $\phi_G$ acquires an additional phase $n_V\pi$, namely, $\phi_G\rightarrow\phi_G+\phi_0+n_V\pi$, where $n_V$ is the total number of vortices in regions II and IV.

\begin{figure}[tbp]
\begin{center}
\includegraphics[width=4in]{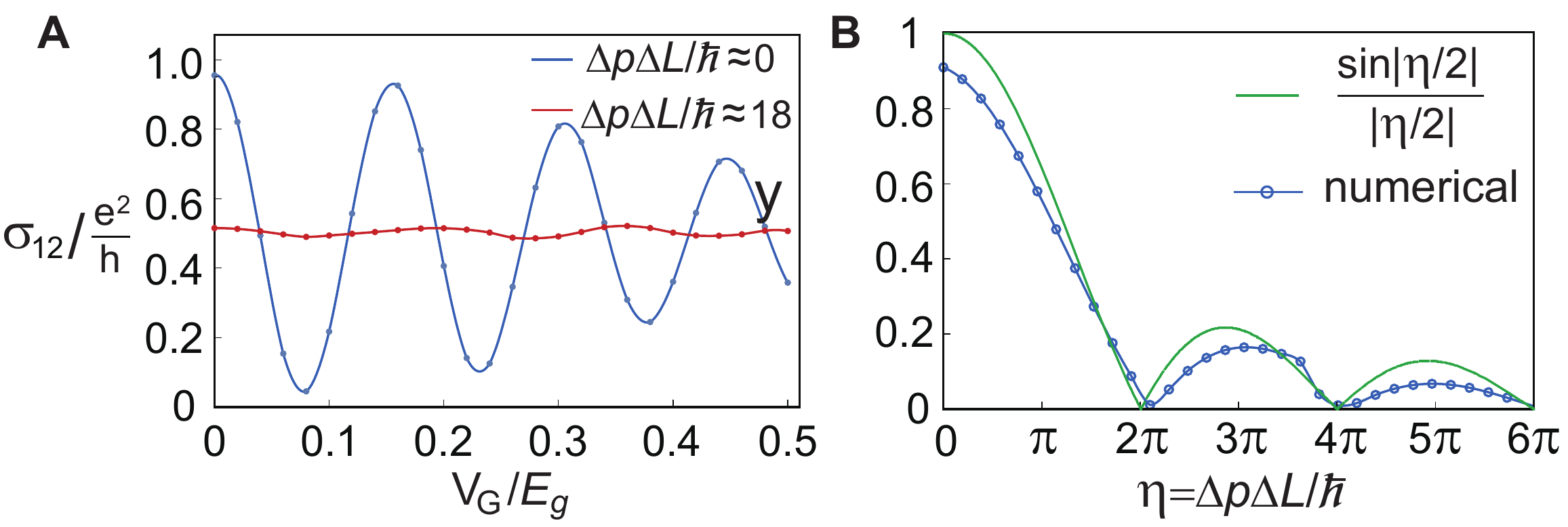}
\end{center}
\caption{Numerically calculated $\sigma_{12}$ oscillation for the Corbino junction. (A) $\sigma_{12}$ calculated for $\Delta p\Delta L/\hbar\approx 0$ and $18$ as a function of $V_G$, respectively. (B) The peak-to-valley amplitude $y$ of $\sigma_{12}$ in units of $e^2/h$ with respect to $\eta=\Delta p\Delta L/\hbar$, which is roughly given by $y=\sin|\eta/2|/|\eta/2|$.}
\label{fig3}
\end{figure}

\section*{Decoherence}

There are mainly two effects contributing to the decoherence of chiral Majorana fermions.
The first is the non-monochromaticity of the incident electron wave packet, which is characterized by a momentum uncertainty $\Delta p\approx 2\pi\hbar/l_W$ for a wave packet of width $l_W$. In general, the (effective) path lengths of the four chiral Majorana modes $\gamma_i$ $(1\le i\le4)$ in Fig. \ref{fig2}A may differ by a length scale $\Delta L$, and the $\sigma_{12}$ oscillation is sharp only if $\Delta p\Delta L<2\pi\hbar$. As a demonstration, we numerically examine the time evolution of an electron wave packet from lead $1$ within an energy window $v_F[-\Delta p/2,\Delta p/2]$ on a lattice model of the Corbino junction and calculate $\sigma_{12}$ \cite{suppl}. Fig. \ref{fig3}A shows $\sigma_{12}$ as a function of $V_G/E_g$ for $\Delta p\Delta L/\hbar\approx 0$ and $18$, respectively, where $E_g$ is the QAHI bulk gap. The modulation of the $\sigma_{12}$ amplitude by $V_G$ is due to the effective change of $\Delta L$ as a result of the change in $v_F$ on the edge covered by voltage gate $V_G$. Fig. \ref{fig3}B shows the peak-to-valley amplitude $y=\Delta\sigma_{12}/(e^2/h)$ as a function of $\eta=\Delta p\Delta L/\hbar$, where we find the amplitude roughly decays as $y=|\sin(\eta/2)/(\eta/2)|$. In the experiments, the temperature $T$ yields a momentum uncertainty $\Delta p\approx k_BT/v_F$, where $k_B$ is the Boltzmann constant. For the Cr-doped (Bi,Sb)$_2$Te$_3$ thin film QAHI with superconducting proximity studied in Ref. \cite{he2017}, the Fermi velocity is of order $\hbar v_F\sim 3$eV$\cdot$\AA \cite{liu2010b}, and the temperature $T$ reaches as low as $20$mK. This requires a path length difference $\Delta L\sim 100\mu$m or smaller, which is experimentally feasible \cite{he2017,fox2017}.

The second effect causing decoherence is the inelastic scattering. The inelastic scattering of charged chiral fermions $\psi_i$ mainly originates from the electron-phonon coupling, %$H_{ep}=\lambda_{ep}\psi_i^\dag\psi_i\nabla\cdot\mathbf{u}$, where $\mathbf{u}$ is the phonon field representing the lattice displacement.
which yields an inelastic scattering length $l_{in}\propto T^{-p/2}$ at temperature $T$ \cite{thouless1977,pruisken1988,huckestein1990}. For integer quantum Hall systems, $l_{in}$ exceeds $10^2\mu$m at $T\sim20$mK \cite{koch1991b}, while $l_{in}$ is expected to be smaller for QAHI \cite{chang2013b}. In contrast, since the electron-phonon coupling is odd under charge conjugation, the neutral chiral Majorana fermions $\gamma_i$ are immune to phonon coupling. Instead, their lowest order local interaction is of the form $\gamma_i\partial_x\gamma_i\partial_x^2\gamma_i\partial_x^3\gamma_i$ \cite{fu2009a}, which is highly irrelevant. Therefore, $l_{in}$ of $\gamma_i$ in TSCs should be much longer than that of $\psi_i$ in QAHIs. If the $\sigma_{12}$ interference is to be observed, the sizes of the QAHI and TSC regions in the junction have to be within their inelastic scattering lengths $l_{in}$, respectively.

\section*{Conclusion}

In summary, we have introduced the appealing possibility of performing topological quantum computations via propagations of 1D chiral Majorana fermion wave packets, which are physically equivalent to the braiding of MZMs. The Corbino junction above gives a minimal demonstration of single-qubit quantum-gate operations with chiral Majorana fermions, and the conductance of the junction provides a natural readout for the final qubit states. Most importantly, this circumvents two main experimental difficulties in quantum computations with MZMs: the braiding operation of MZMs and the readout of the qubit states. The high velocity of chiral Majorana edge modes also makes the quantum gates $10^3$ times faster than those of other quantum computation schemes \cite{haffner2008,gambetta2017}.
Furthermore, the development of single electron source \cite{feve2007} makes the injection and detection of a single electron wave packet qubit on edges possible.
Yet in the current stage we still face difficulties which are also encountered by the MZM quantum computation scheme: the error correction of the phase gate $R_{\phi_G}$ \cite{Bonderson2010,Bravyi2005} and nondemolitional four-Majorana implementation of the controlled NOT gate \cite{Bravyi2005,Bravyi2006,alicea2011}.
If one could overcome these difficulties, one may in principle achieve universal quantum computation using chiral Majorana fermion devices, which would have a high computation speed. Finally, we remark that the conductance oscillation in the Corbino junction, if observed, will also unambiguously prove the existence of quantum coherent chiral Majorana fermions in the experiment \cite{he2017,fu2009a,akhmerov2009,stern2006,bonderson2006,nilsson2010}.

\bibliography{Cor_ref}

\begin{thebibliography}{10}

\bibitem{qi2011}
X.-L. Qi, S.-C. Zhang, {\it Rev. Mod. Phys.\/} {\bf 83}, 1057 (2011).

\bibitem{qi2010b}
X.-L. Qi, T.~L. Hughes, S.-C. Zhang, {\it Phys. Rev. B\/} {\bf 82}, 184516
  (2010).

\bibitem{chung2011}
S.~B. Chung, X.-L. Qi, J.~Maciejko, S.-C. Zhang, {\it Phys. Rev. B\/} {\bf 83},
  100512 (2011).

\bibitem{wang2015a}
J.~Wang, Q.~Zhou, B.~Lian, S.-C. Zhang, {\it Phys. Rev. B\/} {\bf 92}, 064520
  (2015).

\bibitem{strubi2011}
G.~Str\"ubi, W.~Belzig, M.-S. Choi, C.~Bruder, {\it Phys. Rev. Lett.\/} {\bf
  107}, 136403 (2011).

\bibitem{he2017}
Q.~L. He, {\it et~al.\/}, {\it Science\/} {\bf 357}, 294 (2017).

\bibitem{moore1991}
G.~Moore, N.~Read, {\it Nucl. Phys. B\/} {\bf 360}, 362  (1991).

\bibitem{kitaev2006}
A.~Kitaev, {\it Annals of Physics\/} {\bf 321}, 2  (2006). January Special
  Issue.

\bibitem{read2000}
N.~Read, D.~Green, {\it Phys. Rev. B\/} {\bf 61}, 10267 (2000).

\bibitem{kitaev2001}
A.~Y. Kitaev, {\it Physics-Uspekhi\/} {\bf 44}, 131 (2001).

\bibitem{lutchyn2010}
R.~M. Lutchyn, J.~D. Sau, S.~Das~Sarma, {\it Phys. Rev. Lett.\/} {\bf 105},
  077001 (2010).

\bibitem{ivanov2001}
D.~A. Ivanov, {\it Phys. Rev. Lett.\/} {\bf 86}, 268 (2001).

\bibitem{kitaev2003}
A.~Kitaev, {\it Ann. Phys.\/} {\bf 303}, 2 (2003).

\bibitem{alicea2011}
J.~Alicea, Y.~Oreg, G.~Refael, F.~von Oppen, M.~P.~A. Fisher, {\it Nat.
  Phys.\/} {\bf 7}, 412 (2011).

\bibitem{alicea2012}
J.~Alicea, {\it Reports on Progress in Physics\/} {\bf 75}, 076501 (2012).

\bibitem{aasen2016}
D.~Aasen, {\it et~al.\/}, {\it Phys. Rev. X\/} {\bf 6}, 031016 (2016).

\bibitem{karzig2017}
T.~Karzig, {\it et~al.\/}, {\it Phys. Rev. B\/} {\bf 95}, 235305 (2017).

\bibitem{liu2008}
C.-X. Liu, X.-L. Qi, X.~Dai, Z.~Fang, S.-C. Zhang, {\it Phys. Rev. Lett.\/}
  {\bf 101}, 146802 (2008).

\bibitem{yu2010}
R.~Yu, {\it et~al.\/}, {\it Science\/} {\bf 329}, 61 (2010).

\bibitem{chang2013b}
C.-Z. Chang, {\it et~al.\/}, {\it Science\/} {\bf 340}, 167 (2013).

\bibitem{suppl}
See Supplemental Material for details.

\bibitem{bonderson2006}
P.~Bonderson, A.~Kitaev, K.~Shtengel, {\it Phys. Rev. Lett.\/} {\bf 96}, 016803
  (2006).

\bibitem{an2011}
S.~{An}, {\it et~al.\/}, {\it ArXiv e-prints\/} p. 1112.3400 (2011).

\bibitem{liu2010b}
C.-X. Liu, {\it et~al.\/}, {\it Phys. Rev. B\/} {\bf 82}, 045122 (2010).

\bibitem{fox2017}
E.~J. {Fox}, {\it et~al.\/}, {\it ArXiv e-prints\/} p. 1710.01850 (2017).

\bibitem{thouless1977}
D.~J. Thouless, {\it Phys. Rev. Lett.\/} {\bf 39}, 1167 (1977).

\bibitem{pruisken1988}
A.~M.~M. Pruisken, {\it Phys. Rev. Lett.\/} {\bf 61}, 1297 (1988).

\bibitem{huckestein1990}
B.~Huckestein, B.~Kramer, {\it Phys. Rev. Lett.\/} {\bf 64}, 1437 (1990).

\bibitem{koch1991b}
S.~Koch, R.~J. Haug, K.~v. Klitzing, K.~Ploog, {\it Phys. Rev. Lett.\/} {\bf
  67}, 883 (1991).

\bibitem{fu2009a}
L.~Fu, C.~L. Kane, {\it Phys. Rev. Lett.\/} {\bf 102}, 216403 (2009).

\bibitem{haffner2008}
H.~H{\"a}ffner, C.~Roos, R.~Blatt, {\it Physics Reports\/} {\bf 469}, 155
  (2008).

\bibitem{gambetta2017}
J.~M. Gambetta, J.~M. Chow, M.~Steffen, {\it npj Quantum Inf.\/} {\bf 3}, 2
  (2017).

\bibitem{feve2007}
G.~F{\`e}ve, {\it et~al.\/}, {\it Science\/} {\bf 316}, 1169 (2007).

\bibitem{Bonderson2010}
P.~Bonderson, D.~J. Clarke, C.~Nayak, K.~Shtengel, {\it Phys. Rev. Lett.\/}
  {\bf 104}, 180505 (2010).

\bibitem{Bravyi2005}
S.~Bravyi, A.~Kitaev, {\it Phys. Rev. A\/} {\bf 71}, 022316 (2005).

\bibitem{Bravyi2006}
S.~Bravyi, {\it Phys. Rev. A\/} {\bf 73}, 042313 (2006).

\bibitem{akhmerov2009}
A.~R. Akhmerov, J.~Nilsson, C.~W.~J. Beenakker, {\it Phys. Rev. Lett.\/} {\bf
  102}, 216404 (2009).

\bibitem{stern2006}
A.~Stern, B.~I. Halperin, {\it Phys. Rev. Lett.\/} {\bf 96}, 016802 (2006).

\bibitem{nilsson2010}
J.~Nilsson, A.~R. Akhmerov, {\it Phys. Rev. B\/} {\bf 81}, 205110 (2010).

\bibitem{ingo2003}
I.~Peschel, {\it Journal of Physics A: Mathematical and General\/} {\bf 36},
  L205 (2003).

\bibitem{ingo2009}
I.~Peschel, V.~Eisler, {\it Journal of Physics A: Mathematical and
  Theoretical\/} {\bf 42}, 504003 (2009).

\bibitem{entin2008}
O.~Entin-Wohlman, Y.~Imry, A.~Aharony, {\it Phys. Rev. B\/} {\bf 78}, 224510
  (2008).

\end{thebibliography}
\bibliographystyle{Science}

\section*{Acknowledgments}
B.L. acknowledges the support of Princeton Center for Theoretical Science at Princeton University. X.-Q.S. and S.-C.Z. acknowledges support from the US Department of Energy, Office of Basic Energy Sciences
under contract DE-AC02-76SF00515. A.V. acknowledges the Gordon and Betty Moore Foundation's EPiQS Initiative through Grant GBMF4302. X.-L.Q. acknowledges support from David and Lucile Packard Foundation.

\newpage

\section*{Supplementary Material}
The supplementary material is organized as follows. In Sec. 1 we show the 2D lattice Hamiltonians of QAHI and $p+ip$ TSC we use for calculations of entanglement entropy change in the QAHI-TSC-QAHI junction and conductance in the Corbino junction. Sec. 2 gives the details of entanglement entropy numerical calculation for a QAHI-TSC-QAHI junction lattice model during the evolution of an incident electron above the fermi sea. Sec. 3 reviews the generalized Landauer-B\"uttiker formula for two-terminal conductance of a superconducting junction, while Sec. 4 shows the numerical calculation for $\sigma_{12}$ oscillation of a Corbino junction in a 2D lattice as a function of the gate voltage $V_G$. Sec. 5 shows that the nonzero chemical potential on QAHI edges induces a phase shift to $\phi_G$ in the formula of $\sigma_{12}$ in the Corbino junction. Finally, in Sec. 6, we provide a Bloch sphere illustration of the single qubit quantum gate that we propose to implement by the Corbino junction.

\section{ Model Hamiltonian for simulation}

In this section, we present the 2D lattice model Hamiltonian that we will use for later numerical calculations. The structures that we study in the main text consists of a quantum anomalous Hall insulator (QAHI), where we add s-wave superconductivity pairing to induce $p+ip$ chiral topological superconductor (TSC) or add voltage gate to change the chemical potential of edge states. The lattice model Hamiltonian for QAHI we adopt is as follows:
\begin{equation}
H_{QAH}=\sum_{\bf{k}} c_{\bf{k}}^{\dagger}[(A \sin k_x \sigma_x + A \sin k_y  \sigma_y+ (M-B(\cos k_x+\cos k_y))\sigma_z -\mu] c_{\bf{k}},
\label{QAH}
\end{equation}
where $c_{\bf{k}}=(c_{\bf{k}\uparrow}, c_{\bf{k},\downarrow})^{T}$ are fermion operators in momentum space and $\sigma_x$, $\sigma_y$ and $\sigma_z$ are Pauli matrices. We work in the dimensionless unit with lattice constant $a=1$ and set $A=1$, $B=5/2$, $M=4$ and $\mu=0$. The band parameters are chosen such that the the valence band has a non-trivial Chern number and therefore describe a QAHI. In the calculation for the QAHI-TSC-QAHI junction or the Corbino junction, we write the above Hamiltonian in the real space with an open boundary condition at the edges between the junction and the vacuum.

The $p+ip$ TSC is realized by adding an $s$-wave superconductivity pairing $\sum_{\bf{r}}\frac{1}{2} \Delta({\bf{r}})c_{\bf{r}}^{T}i\sigma_{y} c_{\bf{r}}+h.c$ into the Hamiltonian Eq.(\ref{QAH}), where $c_{\bf{r}}=(c_{\bf{r}\uparrow}, c_{\bf{r},\downarrow})^{T}$ are fermion operators in the real space. We choose to set $\Delta({\bf{r}})=\Delta=2$ in the superconducting regions, which drives the regions into a $p+ip$ TSC. We model the static electrical potential induced by voltage gate with a chemical potential term $\sum_{\bf{r}} V({\bf{r}}) c_{\bf{r}}^{\dagger}c_{\bf{r}}$, where $V({\bf{r}})=V_G$ inside the gated region $V({\bf r})=0$ outside. The full model Hamiltonian can be summarized as
\begin{equation}
H=H_{QAH}+\sum_{\bf{r}}\frac{1}{2} [\Delta({\bf{r}})c_{\bf{r}}^{T}i\sigma_{y} c_{\bf{r}}+h.c]+\sum_{\bf{r}} V({\bf{r}}) c_{\bf{r}}^{\dagger}c_{\bf{r}}.
\label{Hamiltonian}
\end{equation}

In all simulations, the model Hamiltonian will be kept at the fixed parameters where $a=1$, $A=1$, $B=5/2$, $M=4$, $\mu=0$ and $\Delta=2$. Several useful quantities are the Fermi velocity $v_F$ of the edge modes, which is equal to $1$ at zero chemical potential. The energy gap is $E_g=2$ for the QAHI regions, and is $1$ in the TSC regions.

\section{Entanglement entropy during the propagation of $\gamma_i$}\label{entanglement}

In this section, we discuss the entanglement entropy change of the QAHI-TSC-QAHI junction during the propagation of an incident electron from lead $1$. In the case of the Majorana zero mode(MZM), if one splits a system into two subsystems $A$ and its complement $A^{c}$, the braiding of one MZM in subsystem $A$ with another MZM in subsystem $A^{c}$ creates an entanglement entropy $\log 2$ for the subsystem $A$. This is also expected to be true in our case of propagation of chiral Majorana fermion wave packets. Indeed, a nonvanishing increment in the value of entanglement entropy is a generic signature of non-Abelian transformations (gate operations).
%, as the initial state will mix with other degenerate states upon the braiding operations, and as a result, the entanglement entropy associated with an entanglement cut that segregates the pair of anyons undergoing braiding will increase afterwards.
%On the contrary, for Abelian states, the entanglement entropy remains unchanged upon braiding of anyons. Thus, the evolution of entanglement entropy throughout the braiding operation is a powerful tool that can easily distinguish non-Abelian from Abelian statistics.

\begin{figure}[htbp]
\centering
\includegraphics[width=0.5\linewidth]{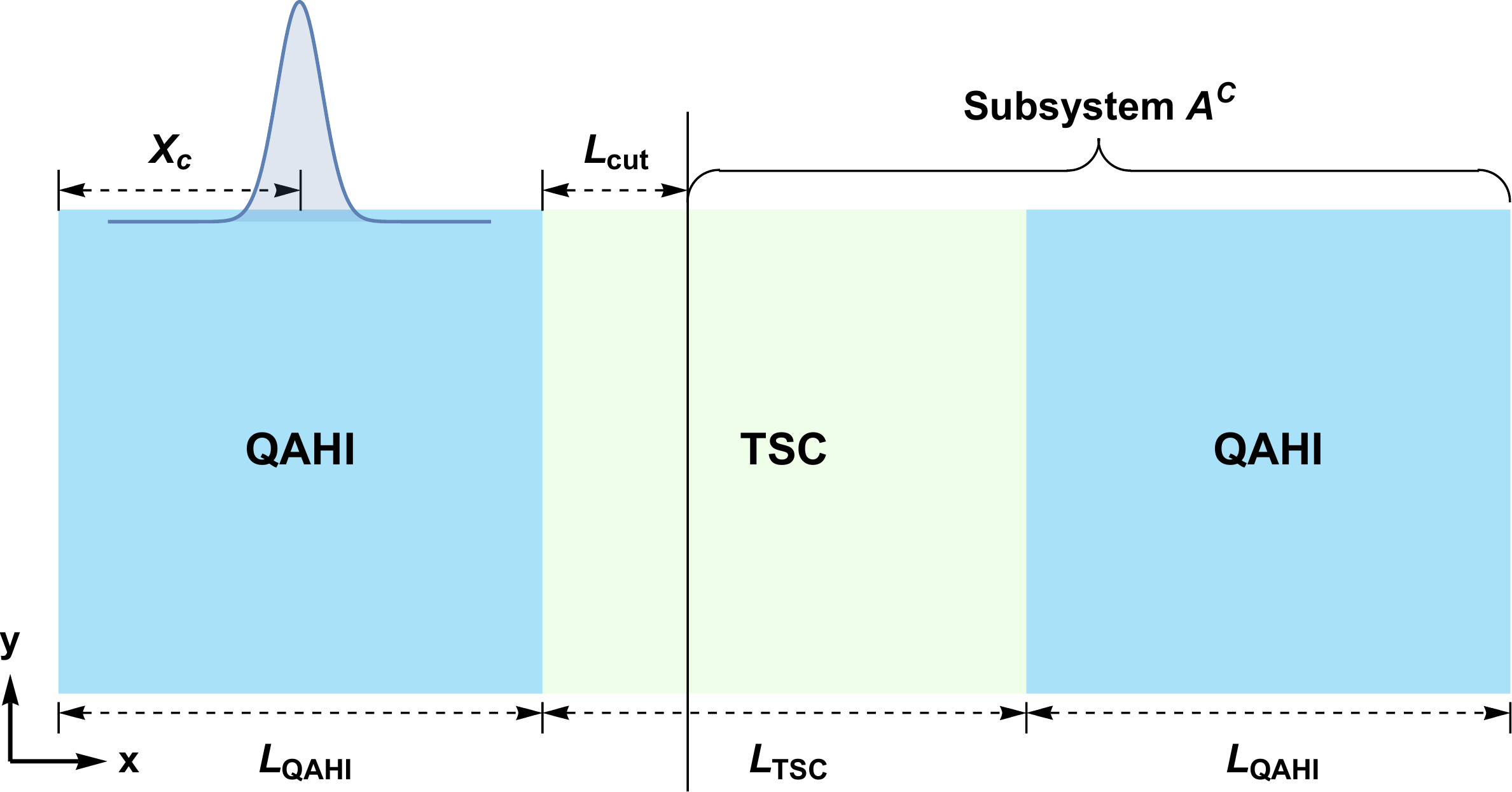}
\caption{The geometry of a QAHI/TSC/QAHI junction. We align the QAHI regime, TSC regime and the other QAHI regime in the x direction. The length of each QAHI regime in x direction is $L_{QAHI}$ while the length of TSC regime in x direction is $L_{TSC}$. A cut along y direction at the TSC regime is made at a distance $L_{cut}$ to the boundary of TSC and the left QAHI. We define subsystem $A$ as the subsystem to the left of the cut and we denote its compliment $A^{C}$. For illustration purpose, we only label the compliment subsystem $A^{C}$. The position of the initial wave packet is centered at a distance $X_{c}$ to the boundary of vacuum and the left QAHI. The simulation of Fig. 1E in the main text is run at the geometry parameters $L_{QAHI}=30$, $L_{TSC}=20$, $X_{c}=10$ and $L_{cut}=10$.}
\label{figS1}
\end{figure}

We design the Hamiltonian defined in Eq.(\ref{Hamiltonian}) for a QAHI-TSC-QAHI junction on a lattice as shown in Fig. S1. The length of each QAHI region in $x$ direction is $L_{QAHI}$ while the length of TSC region in x direction is $L_{TSC}$. A cut along $y$ direction in the TSC region is made at a distance $L_{cut}$ to the boundary of TSC and the left QAHI. We define subsystem $A$ as the subsystem to the left of the cut and we denote its compliment $A^{C}$ in Fig. \ref{figS1}. The entanglement entropy of subsystem $A$ is given by
\begin{equation}
S_{E}=-Tr (\rho_{A} \log \rho_{A})\ ,
\end{equation}
where $\rho_{A}$ is the reduced density matrix of the quantum state of subsystem A. With the BdG Hamiltonian adopted, the system consists of non-interacting fermionic quasiparticles. We denote the annihilation operators of the BdG quasiparticle eigenstates as $\alpha_{m}$, $m=1,...,n$. The many-particle state for the fermi sea of the system is then $|0\rangle$ satisfying $\alpha_m|0\rangle=0$.

We then consider the evolution of an electron wave packet state injected from lead $1$, given by $|\Psi(t)\rangle=\beta^\dag(t)|0\rangle$, where $\beta^\dag(0)$ is a chosen creation operator of an electron wave packet at time $t=0$ located near lead $1$ on the QAHI edge, and $\beta^\dag(t)=e^{iHt}\beta^\dag(0)e^{-iHt}$ is its time evolution. The wave packet is restricted within an energy window $[0,v_{F}\Delta p]$, which is smaller than the minimal bulk gap of the system.

The entanglement entropy of the noninteracting fermion states (i.e., Slater determinant states) $|\Psi(t)\rangle$ and $|0\rangle$ are given by \cite{ingo2003,ingo2009}
%At $t=0$, We initialize a wave packet composed from quasiparticle states with momentum uncertainty $\Delta p$ (namely, those quasiparticle states have energy in the window of $[0,v_{F}\Delta p]$) centered on the upper edge at a distance $X_{c}$ to the boundary of vacuum and the left QAHI. We denote the operator to annihilate such a single quasiparticle state operator as $\alpha_{1}'(0)$ which is a superposition of quasiparticle states $\alpha_{i}$, $i=1,..n$. At time t, this operator evolves to $\alpha_{1}'(t)$ in Heisenberg picture. We can find an orthogonal basis of the linear space spanned by $\alpha_{1}, \alpha_{2},...., \alpha_{n}$ as $\alpha_{1}'(t), \alpha_{2}'(t),..., \alpha_{n}'(t)$. In this basis, only state annihilated by  $\alpha_{1}'(t)$ is occupied while other states annihilated by $\alpha_{i}'(t)$ are unoccupied. The entanglement entropy of such wave function is greatly simplified\cite{ingo2003,ingo2009}:
\begin{equation}
S_{E}(t)=-\sum_{\alpha} C_{\alpha}(t) \log C_{\alpha}(t)\ ,\quad S_{E0}=-\sum_{\alpha} C^0_{\alpha} \log C^0_{\alpha},
\label{SA}
\end{equation}
respectively, where $C_{\alpha}(t)$ and $C_\alpha^0$ are eigenvalues of the correlation matrices defined as follows:
\[C_{is,js'}(t)=
\left(
\begin{array}{cc}
\langle \Psi(t)|c_{i s} c_{j s'}^{\dagger} |\Psi(t)\rangle & \langle\Psi(t)| c_{i s} c_{j s'} |\Psi(t)\rangle \\
\langle\Psi(t)| c_{i s}^{\dagger} c_{j s'}^{\dagger}|\Psi(t) \rangle & \langle\Psi(t)| c_{i s}^{\dagger} c_{j s'} |\Psi(t) \rangle
\end{array}
\right),\]
\begin{equation}
C_{is,js'}^0=
\left(
\begin{array}{cc}
\langle 0|c_{i s} c_{j s'}^{\dagger} |0\rangle & \langle0| c_{i s} c_{j s'} |0\rangle \\
\langle0| c_{i s}^{\dagger} c_{j s'}^{\dagger}|0 \rangle & \langle0| c_{i s}^{\dagger} c_{j s'} |0 \rangle \end{array}
\right).
\label{correlation}
\end{equation}
Here $c_{is}$ is the electron annihilation operator on site $i$ in the subsystem A, while $s$, $s'$ are the spin indices. The correlation matrix $C^0$ of the fermi sea can be calculated from the eigenstate operators $\alpha_m$. Once the commutators of $c_{is},c_{is}^\dag$ with the $\beta(t)$, $\beta^\dag(t)$ are determined, the correlation matrix $C(t)$ of the wave packet state can be calculated based on $C^0$, and the entanglement entropy can be calculated numerically.

We calculate the time evolution of the entanglement entropy $S_E(t)-S_{E0}$ using geometry parameters $L_{QAHI}=30$, $L_{TSC}=20$, $X_{c}=10$ and $L_{cut}=10$. We set the wave packet to contain quasiparticle states in an energy window $[0,0.75]$. The wave packet is created by projecting an electron wave packet onto the quasiparticle states in this energy window. Summary of the geometry parameters is given in Fig. \ref{figS1}, and the evolution of the entanglement entropy is plotted in Fig. 1E of the main text. We can clearly that after $t=60$ when the wave packet has left the TSC regime, the entanglement entropy increase of subsystem A is quantized at $\log 2$.

\section{ Calculation of the two terminal conductance}\label{conductance}

In this supplementary section, we briefly review the calculation of the two terminal conductance for the Corbino junction. The two terminal conductivity from the lead 1 to the lead 2 can be obtained from the generalized Landauer-Buttiker formula\cite{entin2008}:
\begin{equation}
I_{i}=\frac{e^2}{h}[(1-R^{i}+R_{A}^{i})(V_{i}-V_{SC})-\sum_{j\neq i}(T^{ji}-T_{A}^{ji})(V_{j}-V_{SC})], \quad i=1,2,
\label{Landauer}
\end{equation}
where $I_{i}$ is the current flowing out of the lead $i$, $V_{i}$ is the voltage of the lead $i$, and $T_{ij}$, $T_{A}^{ij}$ are the normal transmission and Andreev transmission probabilities from leads $i$ to $j$ ($j\neq i$), while $R^{i}$ and $R_{A}^{i}$ are the normal reflection and Andreev reflection from the lead $i$ back to itself, respectively. As a consistency check, the conductance $\sigma_{12}$ of the Corbino junction calculated this way should agree with our prediction in the main text based on chiral Majorana fermion propagations.

We simulate the time evolution of an electron wave packet initialized inside the lead 1 region using the Hamiltonian from Eq. (\ref{Hamiltonian}). At the time when the wave packet reflects (transmits) to the lead 1 (lead 2) neighbourhoods, we stop the time evolution and compute the probability of reflection and transmission, namely $T_{ij}$, $T_{A}^{ij}$, $R^{i}$ and $R_{A}^{i}$, from the wave function. Note that if we connect the electron source directly across leads 1 and 2, we also have an additional constrain:
\begin{equation}
I_{1}=-I_{2}=I.
\label{constrain}
\end{equation}
From Eq. (\ref{Landauer}) and Eq. (\ref{constrain}), we can then solve for two terminal conductivity $\sigma_{12}=(V_{1}-V_{2})/I$.

\section{Decoherence effect from non-monochromaticity}\label{wavepacket}
In the main text, we have discussed the decoherence effect from the non-monochromaticity of the incident electron wave packet.  The non-monochromaticity is described by the momentum uncertainty $\Delta p$ of the electron wave packet together with a length scale characterizes the length difference $\Delta L$ of the four chiral Majorana modes $\gamma_{i}$ ($1\le i \le 4 $). In this section, we shall discuss the precise definition of these parameters in simulation and the method to study the dependence of oscillation amplitude on them.

\begin{figure}[htbp]
\centering
\includegraphics[width=0.6\textwidth]{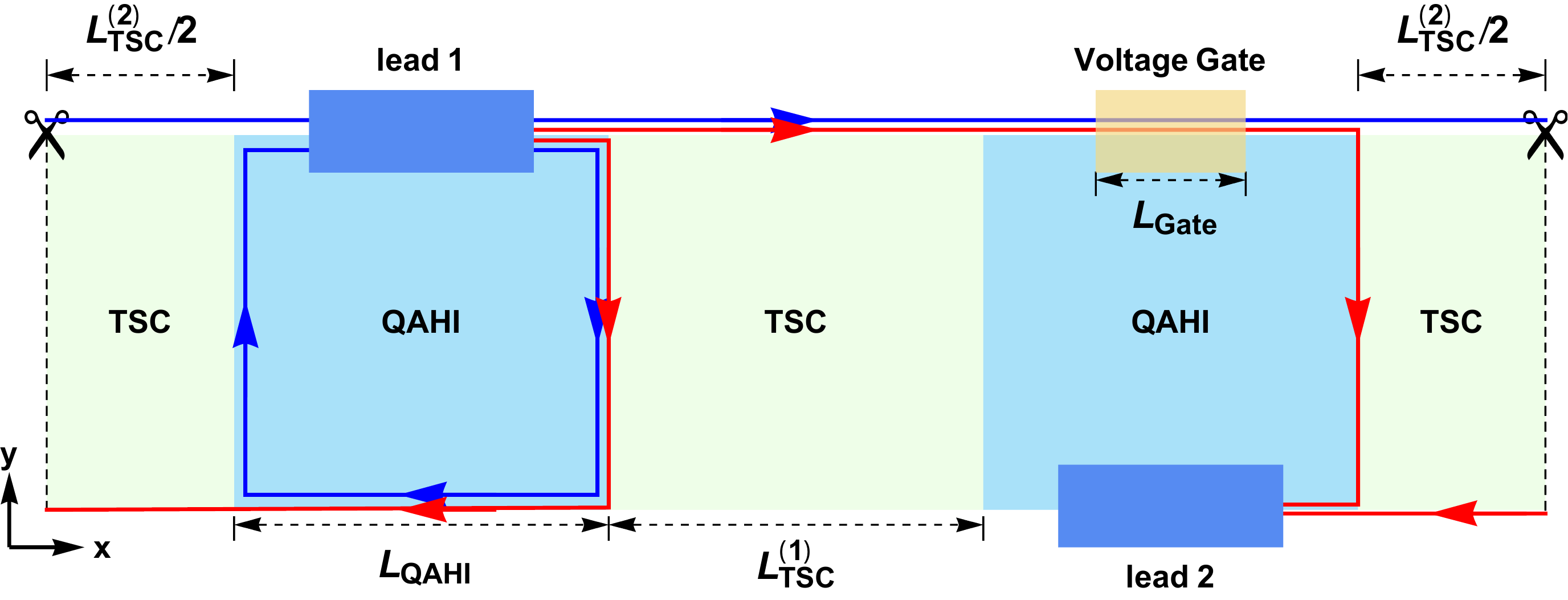}
\caption{Illustration of chiral Majorana interferometry: A band of QAHI with two TSC regimes induced by proximity to a s-wave superconductor. The lengths of of QAH regimes, TSC regimes and the voltage gate regime are denoted as $L_{QAHI}$, $L_{TSC}^{(1)}$, $L_{TSC}^{(2)}$ and $L_{Gate}$, respectively. If one consider an incident electron wave packet from the lead 1, we can decompose it into a superposition of two Majorana fermions. Two red lines are paths for those Majorana fermions to travel from the lead 1 to the lead 2 while two blue lines are paths for those Majorana fermions to travel back to the lead 1. The probability for a charge from the lead 1 to transmit/reflect is contributed by the red/blue paths. The path difference of two transmitted/reflected paths from the lead 1 is $\Delta L=|L_{TSC}^{(1)}-L_{TSC}^{(2)}|$. }
\label{figS2}
\end{figure}

As shown in Fig. \ref{figS2}, we put the Corbino junction on a cylindrical lattice with left and right vertical dashed lines identified, which is equivalent to the Corbino geometry. We can consider an incident electron wave packet from the lead 1. In simulation, we obtain a wave packet of momentum uncertainty $\Delta p$ in the following way. We initialize an electron wave packet broader than $\hbar/\Delta p$. Then we project this wave function onto the energy eigenspace of Hamiltonian from Eq.(\ref{Hamiltonian}) in the energy window $v_{F}[-\Delta p/2, +\Delta p/2]$ and normalize the projected wave function as $\psi(0)$. We shall define $\psi(0)$ as the initial electron wave packet with momentum uncertainty $\Delta p$. Notice that this initial condition is slightly different from the calculation for entanglement entropy in section \ref{entanglement} because the negative energy state represents a hole of quasiparticle which is impossible to generate from ground state with no quasiparticles at zero temperature. Here we are considering the non-monochromaticity of electron wave packet from the finite temperature effect and this initial condition is physical.

A suitable perspect is to consider the electron wave packet as a superposition of wave packets of two Majorana fermions. Upon time evolution, the fate of the two Majorana fermions is either recombination to a particle/hole at the lead 1 or at the lead 2. For the process that the wave packet ends up back at the lead 1, the probability is contributed by two paths shown as two blue lines in Fig. \ref{figS2}. In a precise fashion, this can be interpretted as a interferometry of chiral Majorana fermions: the electron wave packet passes through a "beam splitter" , travels through two arms as through the chiral Majorana mode and recombines at the lead 1. The length difference of the two arms of the interferometry is $\Delta L^{(1)}=|L_{TSC}^{(1)}+L_{TSC}^{(2)}-2L_y|$ and we can expect the interference effect in the probability of propagating back to be measurable when $\Delta L^{(1)} \Delta p<h$. For the process that the wave packet transmits to the lead 2, similarly, the probability is contributed by two paths shown as two red lines in Fig. \ref{figS2}. The length difference of the two paths is $\Delta L^{(2)}=|L_{TSC}^{(1)}-L_{TSC}^{(2)}|$ and the condition for the interference is $\Delta L^{(2)} \Delta p<h$. For illustration purpose, we study the case when $L_{y}=L_{TSC}^{(1)}$ so that $\Delta L^{(1)}=\Delta L^{(2)}=\Delta L$ so that a unique length scale $\Delta L$ is defined.

In simulation, we fix the geometry parameters at $L_{TSC}^{(1)}=L_{y}=20$, $L_{QAH}=30$ and $L_{Gate}=20$ and vary $\Delta L=L_{TSC}^{(2)}-L_{TSC}^{(1)}$ from 0 to 30. For each $L_{TSC}^{(2)}$, we initialize a wave packet at the lead 1 region with momentum uncertainty $\Delta p/v_{F}\hbar=0.6$. We can simulate the time evolution of the wave packet and obtain $\sigma_{12}$ as described in the previous section for $V_{G}$ from 0 to 1. At $\Delta L=0$ $(\Delta L \Delta p/\hbar=18)$ and $\Delta L=30$ $(\Delta L \Delta p/\hbar=18)$, the dependence of $\sigma_{12}$ on $V_{G}$ is shown in Fig. 3A in the main text with an oscillation feature. We can also observe similar oscillation for other $\Delta L$ and the peak-to-valley oscillation amplitude has a dependence on $\Delta L \Delta p/\hbar$ shown in Fig. 3A in the main text.

\section{Phase shift of $\phi_G$ due to nonzero chemical potential $\mu$}
In this section we discuss the phase shift of $\phi_G$ in the two terminal conductance $\sigma_{12}$ of the Corbino junction due to chemical potential and static disorders on the QAHI edges.
When the chemical potential $\mu$ on a QAHI edge is nonzero, the Hamiltonian of the corresponding charged chiral edge state $\psi$ is
\begin{equation}
H_{F}(x)=-i\hbar v_F\psi^\dag(x)\partial_{x}\psi(x)-\mu(x)\psi^\dag(x)\psi(x)\ .
\end{equation}
Solving the Shr\"odinger equation yields an electron wave function
\begin{equation}
\psi(x,t)=\exp\left[\frac{i}{\hbar v_F}\int_0^x\mu(x')dx'\right]\varphi_0(x-v_Ft)\ ,
\end{equation}
where $\varphi_0(x)$ is an arbitrary function of $x$. Therefore, a chiral fermion wave packet accumulates a phase $\phi=\int_{x_1}^{x_2}\mu(x)dx$ after propagation from $x_1$ to $x_2$ which is fixed by the function of chemical potential $\mu(x)$. In contrast, a chiral Majorana fermion always has zero chemical potential as ensured by the particle-hole symmetry of TSC.

In the Corbino junction as shown in Fig. 2A of the main text, assume charged chiral state $\psi_\alpha$ ($\alpha=A,B,C,D$) accumulates an additional chemical potential induced phase $\phi_\alpha$ during propagation on the corresponding QAHI edge. In the odd fermion parity subspace $\{|0_A1_B\rangle,|1_A0_B\rangle\}$, the total unitary transformation becomes
\begin{equation}
|\psi_f\rangle=\left(\begin{array}{cc}e^{-i\phi_B}&0\\0&e^{-i\phi_A}\end{array}\right) ZHR_{\phi_G} \left(\begin{array}{cc}e^{-i\phi_D}&0\\0&e^{-i\phi_C}\end{array}\right)ZH|\psi_i\rangle\ ,
\end{equation}
which is equivalent to insertion of two additional phase gates. As a result, an initial state $|\psi_i\rangle=|1_A0_B\rangle$ transforms into a final state
\begin{equation}
|\psi_f\rangle=e^{-i(\phi_G+\phi_D+\phi_C)/2}\left(e^{-i\phi_B}\cos\frac{\phi_G+\phi_0}{2}|0_A1_B\rangle+ i e^{-i\phi_A}\sin\frac{\phi_G+\phi_0}{2}|1_A0_B\rangle\right)\ ,
\end{equation}
where $\phi_0=\phi_D-\phi_C$. Therefore, the conductance $\sigma_{12}$ becomes
\begin{equation}
\sigma_{12}=(1-|\langle\psi_f|\psi_i\rangle|^2)\frac{e^2}{h}=\frac{1+\cos(\phi_G+\phi_0)}{2}\frac{e^2}{h}\ .
\end{equation}

\section{ Bloch sphere illustration of the Corbino junction}
\begin{figure}[htbp]
\centering
\includegraphics[width=\textwidth]{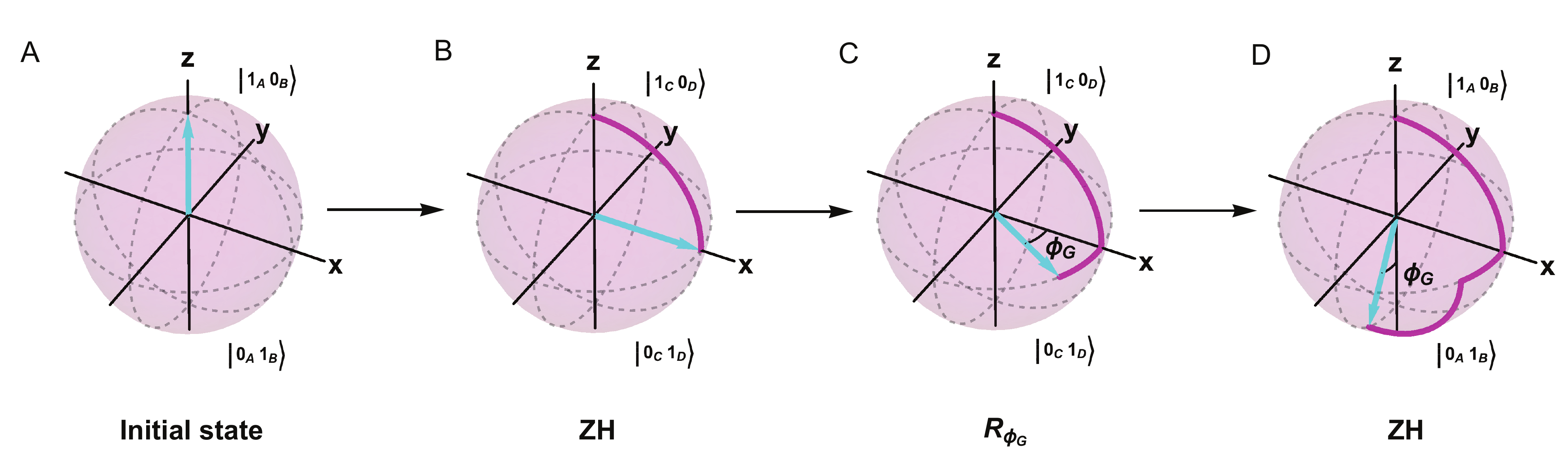}
\label{rotation}
\caption{{\bf{The time evolution of a qubit. }}  ({\bf A}). The electron wave packet is ejected from the lead 1 and occupies one state of $\psi_{A}$ fermion. The qubit at this time is initialized at $|1_{A} 0_{B}\rangle$.  ({\bf B}). The effect of the QAHI I--TSC II--QAHI III junction is a rotation of $\pi/2$ along y axis on the Bloch sphere if we define the qubit state $(|0\rangle, |1\rangle)$ as $(|0_{A} 1_{B} \rangle, |1_{A} 0_{B} \rangle)$ before the wave packet approaches the junction and $(|0_{C} 1_{D} \rangle, |1_{C} 0_{D} \rangle)$ after the wave packet leaves the junction. ({\bf C}). The effect of the voltage gate is a rotation of $-\phi_{G}$ along z axis state on the Bloch sphere of qubit $(|0_{C} 1_{D} \rangle, |1_{C} 0_{D} \rangle)$. ({\bf D}). The effect of the QAHI III--TSC IV--QAHI I junction is a rotation of $\pi/2$ along y axis on the Bloch sphere if we define the qubit state $(|0\rangle, |1\rangle)$ as $(|0_{C} 1_{D} \rangle, |1_{C} 0_{D} \rangle)$ before the wave packet approaches the junction and $(|0_{A} 1_{B} \rangle, |1_{A} 0_{B} \rangle)$ after the wave packet leaves the junction. }
\label{figS3}
\end{figure}

In this section, we present an illustration for the time evolution of the qubit on its Bloch sphere after injecting an electron wave packet from lead 1.  As shown in Fig. 2 in the main text, the charged chiral fermion modes on the QAHI edges are labeled as $\psi_{A}$, $\psi_{B}$, $\psi_{C}$ and $\psi_{D}$. If we regard the charged chiral edge modes of QAHI region I ($\psi_A$ and $\psi_C$) as the data qubit, and those of QAHI region III ($\psi_B$ and $\psi_D$) as the ancilla qubit, the junction can be viewed as a series of quantum gates as shown in Fig. 2B in the main text, with a total unitary evolution $|\psi_f\rangle=ZHR_{\phi_G}ZH|\psi_i\rangle$. The initial state of the wave packet is $|1_{A}0_{B}\rangle$ occupying a $\psi_{A}$ fermion state. The electron wave packet will then approach the TSC II region and leave this region as chiral fermion mode $\psi_{C}$ or $\psi_{D}$. If we define the qubit state $(|0\rangle, |1\rangle)$ as $(|0_{A} 1_{B} \rangle, |1_{A} 0_{B} \rangle)$ before the wave packet approaches the TSC II region and $(|0_{C} 1_{D} \rangle, |1_{C} 0_{D} \rangle)$ after the wave packet leaves the TSC II region, the time evolution of such a process can be viewed as the operator $ZH$ acting on a qubit which is initialized at $|1\rangle$ state at north pole of its Bloch sphere. The $ZH$ operator is a rotation of $\pi/2$ along $y$ axis and upon the operation, the qubit rotates to $+x$ direction on the Bloch sphere. After leaving the TSC II region, the wave packet may enter the voltage gate and the effect of voltage gate is to contribute additional phase $\phi_{G}$ to state $|0_{C} 1_{D}\rangle$ while $0$ to state $|1_{C} D_{D}\rangle$ and therefore is a rotation of $-\phi_{G}$ along z axis in the Bloch sphere of qubit $(|0_{C} 1_{D}\rangle, |1_{C} 0_{D}\rangle)$. Before reaching leads, the wave packet must also approach the TSC IV region and leave this region as chiral fermion mode $\psi_{A}$ or $\psi_{B}$.  The time evolution of such a process can be viewed as the operator $ZH$ rotating the qubit by $\pi/2$ along y axis on the Bloch sphere if we define the qubit state $(|0\rangle, |1\rangle)$ in as $(|0_{C} 1_{D} \rangle, |1_{C} 0_{D} \rangle)$ before the wave packet approaches the junction and $(|0_{A} 1_{B} \rangle, |1_{A} 0_{B} \rangle)$ after the wave packet leaves the junction. From Fig. \ref{figS3}(A-D), we can clearly see the time evolution of the qubit on the Bloch sphere of the process that we have described in this paragraph and the final state at polar angle $\pi-\phi_{G}$ and azimuthal angle $\pi/2$ on the Bloch sphere.  This is an illustrative derivation of Eq. (4) in the main text.

\section{Understanding of the unitary transformation via vortex operators}
Hereby we show the propagation of chiral Majorana wave packets on the TSC edges are physically equivalent to the non-Abelian braiding of $\pi$-flux vortices (which trap MZMs) in the TSC bulk.

The chiral TSC edge is known to be described by the chiral Ising conformal field theory (CFT). By defining $z=v_F^{-1}x-(t-i\delta)$ and $\bar{z}=v_F^{-1}x+(t-i\delta)$, the edge action takes the form
\[S=\int dx dt\gamma(x,t)\bar{\partial}\gamma(x,t),\]
where we use $\partial=\partial_z$ and $\bar{\partial}=\partial_{\bar{z}}$ for short. For imaginary time $t=-i\tau$, the above $z$ and $\bar{z}$ are simply the holomorphic and antiholomorphic coordinates in the $x,\tau$ plane. The equation of motion then indicates $\gamma(x,t)=\gamma(z)$. In addition, the chiral Ising CFT contains the chiral vortex operator $\sigma(x,t)=\sigma(z)$, while $\gamma$ and $\sigma$ satisfy the Ising fusion rules
\begin{equation}
\sigma\times\sigma=1+\gamma,\qquad \gamma\times\gamma=1,\qquad\sigma\times\gamma=\sigma.
\end{equation}
In particular, two $\sigma$ fields may fuse into either a bosonic or a fermionic field, thus $\sigma$ is said to be non-Abelian.

To get a better understanding of the vortex operator $\sigma$, we first recall the nonchiral Ising CFT with action
\[S_{\mathrm{nonchiral}}=\int dx dt[\gamma(z)\bar{\partial}\gamma(z)+\bar{\gamma}(\bar{z})\partial\bar{\gamma}(\bar{z})],\]
which describes the critical point of the $1+1$D transverse field Ising model, where $\gamma(z)$ and $\bar{\gamma}(\bar{z})$ are the right and left moving Majorana fermion fields, respectively. The nonchiral vortex operator is simply the Ising spin $s_z(x,t)=\sigma(z)\bar{\sigma}(\bar{z})$, which is the product of the holomorphic vortex $\sigma(z)$ and the antiholomorphic vortex $\bar{\sigma}(\bar{z})$. When we recover the lattice Ising model defined on sites $x=na$ where $n$ is integral, the right-moving and left-moving Majorana fields $\gamma$ and $\bar{\gamma}$ are well-defined at low energies, and the Ising spin can be expressed in terms of the Majorana fermion fields via a Jordan-Wigner transformation
\begin{equation}
s_z(x,t)=\left[\prod_{x'<x}i\bar{\gamma}(x',t)\gamma(x',t)\right]\left[\gamma(x,t)+\bar{\gamma}(x,t)\right],
\end{equation}
where $i\bar{\gamma}(x',t)\gamma(x',t)$ gives the fermion parity of site $x'$ at low energies.
Therefore, one can roughly decompose it into the product of the following holomorphic and antiholomorphic chiral vortex fields:
\begin{equation}
\sigma(x,t)=\prod_{x'\le x}\gamma(x',t)\ ,\qquad \bar{\sigma}(x,t)=\prod_{x'\le x}\bar{\gamma}(x',t)\ .
\end{equation}
In this way, the chiral vortex fields $\sigma(z)$ and $\bar{\sigma}{(\bar{z})}$ can be understood as half-infinite strings of chiral Majorana fields $\gamma$ and $\bar{\gamma}$ in the interval $[-\infty,x]$ at time $t$, respectively. The chiral Ising CFT fusion rule is then easy to understand in the lattice picture: when the lattice difference $|x_1-x_2|\rightarrow0$, the operator product $\sigma(x_1,t)\sigma(x_2,t)=\prod_{x_1<x\le x_2}\gamma(x,t)$, which is either bosonic or fermionic depending on $(x_2-x_1)/a$ is even or odd. Furthermore, when a Majorana fermion $\gamma$ is moved around a vortex field $\sigma$ in the complex $z$ plane, it necessarily crosses the Majorana string (exchange with a Majorana field on the string) once, and acquires a sign change. Therefore, $\sigma$ behaves as a $\pi$ flux vortex in the complex $z$ plane for $\gamma$.

In the setup of our main text Fig. 1A, the complex chiral fermion $\psi_A$ on the lower left QAH edge is equivalent to two copies of the chiral Ising CFT with the same chirality, namely, one can define two chiral Majorana fields $\gamma_1$ and $\gamma_2$ satisfying $\psi_A=\gamma_1+i\gamma_2$. Accordingly, their vortex fields $\sigma_1$ and $\sigma_2$ can be understood as as half-infinite strings of $\gamma_1$ and $\gamma_2$, respectively (we do not need to worry about boundary conditions since all edges in our setup are open and connected to metallic leads). The incident qubit A, defined by the occupation number of an electron wave packet at $x$ on edge A and at time $t$, is then equivalent to the insertion of two vortices fields, one $\sigma_1(x,t)$ and one $\sigma_2(x,t)$, which together spans a 2D Hilbert space. More explicitly, the operator product $\lim_{x\rightarrow y}\sigma_1(x,t)\sigma_2(x',t)$ tends to $(-1)^{n_F}\gamma_1(x),\ (-1)^{n_F}$ or $(-1)^{n_F}\gamma_2(x)$ depending on $(x-x')/a=1,\ 0$ or $-1$, respectively, where $n_F=\int_{-\infty}^{x}i\gamma_1(x')\gamma_2(x')dx'$ is the fermion number on the left of $x$. Therefore, $\sigma_1$ and $\sigma_2$ fuses into a 2D Hilbert space spanned by two local Majorana operators (wave packets) $\gamma_1(x)$ and $\gamma_2(x)$. The injection of an electron at lead $1$ (state $|1_A\rangle$) corresponds to injection of $\sigma_1$ and $\sigma_2$ in the fermionic fusion channel, and the injection of "nothing" (state $|0_A\rangle$) is the insertion of $\sigma_1$ and $\sigma_2$ in the bosonic fusion channel. We note that since the injected electron wave packet state is a charge eigenstate (i.e., carrying a definite charge), it can only be split into one $\sigma_1$ and one $\sigma_2$, instead of two $\sigma_1$ (or two $\sigma_2$) vortices. Two $\sigma_1$ fields will fuse into a Bogoliubov fermion state which is not a charge eigenstate.

\begin{figure}[htbp]
\centering
\includegraphics[width=0.6\textwidth]{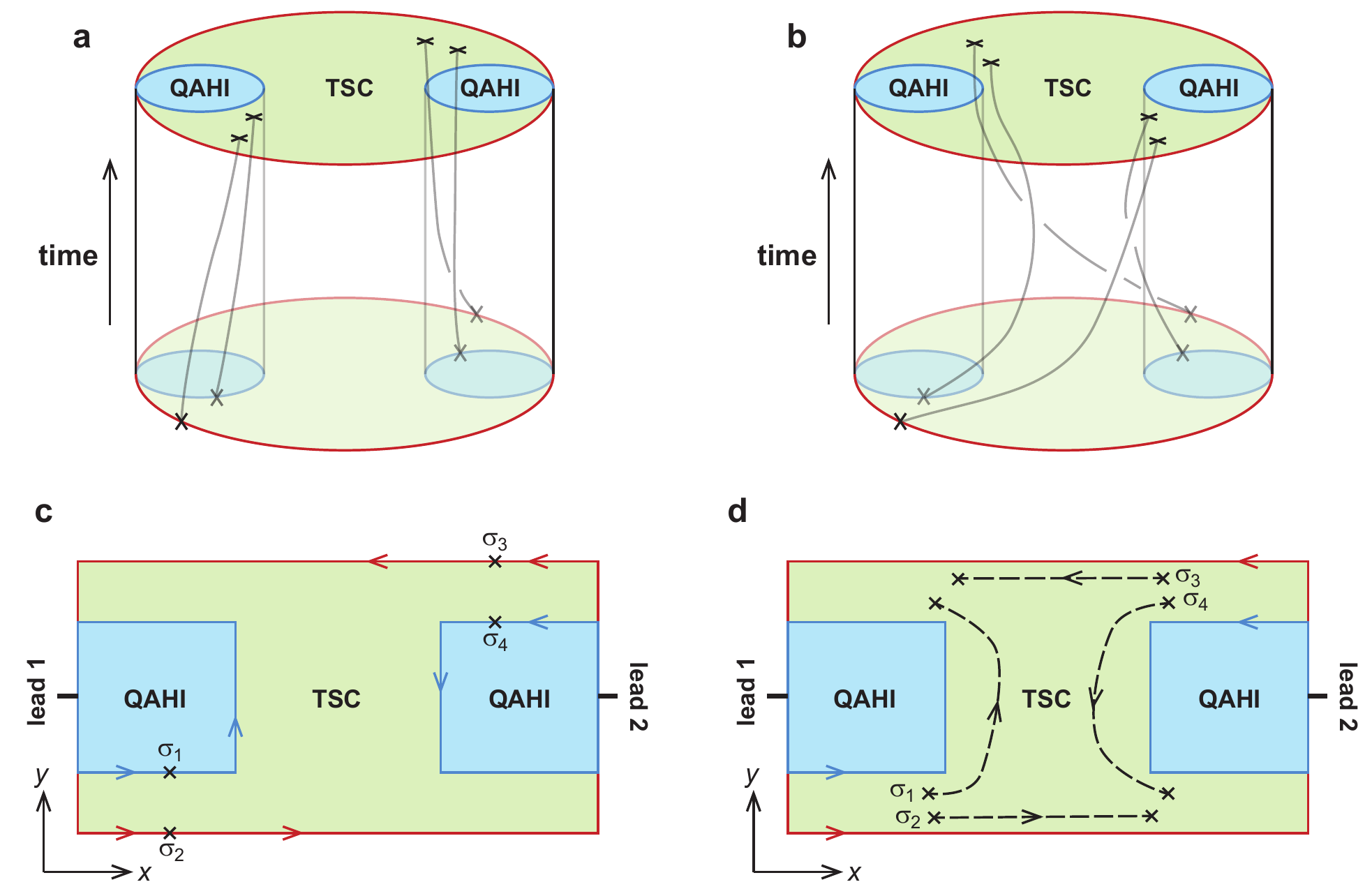}
\label{bulk}
\caption{{\bf{Equivalence between propagation of edge chiral Majorana fermions and bulk braiding of vortices.}} {\bf (A)} The two qubits A and B are equivalent to four vortex operators $\sigma_i$ on the boundary of the TSC at a particular time $t$. Each vortex on the boundary can be connected with a bulk vortex via a Wilson loop. {\bf (B)} The Wilson loop configuration after a bulk vortex braiding and fusion, which is equivalent to that after a boundary evolution as we described in the main text. {\bf (C)} The incident states at edges A and B can be viewed as insertion of four vortices $\sigma_i$. {\bf (D)} By dragging the four vortices into the TSC bulk (along with the incident fermions they trap), and then braid and fuse the vortices in the bulk, one obtain the same final states as that obtained after propagation of chiral Majorana wave packets on the edges.}
\label{figS4}
\end{figure}

We now show that the propagation of chiral Majorana wave packets on the edges is physically equivalent to the braiding/fusion of $\pi$-flux vortices in the bulk of the TSC, which is extensively studied in the literature. As shown in Fig. \ref{figS4}A, in the $2+1$D spacetime of the device, a vortex $\sigma$ inserted at time $t$ on the TSC boundary can be adiabatically connected with a $\pi$ flux vortex in the TSC bulk via a Wilson loop. (In fact, a vortex on the boundary has to continue into the bulk as a Wilson loop to be a legal object in the bulk topological field theory, and the Wilson loop is nothing but the world line of the vortex.) In the topological quantum field theory (TQFT) description of the bulk TSC, all the physical processes are determined by the configuration of Wilson loops in the spacetime. In particular, given two Wilson loops connecting two bulk vortices and two boundary vortices, exchange of two vortices on the boundary $(t,x)$ sheet or braiding of two vortices in the bulk $(x,y)$ plane lead to the same change of the Wilson loop configuration, so they are physically equivalent. Similarly in our case, the creation and fusion of four vortices on the TSC boundary (Fig. \ref{figS4}C, with bulk doing nothing) is equivalent to creation and fusion of four vortices in the TSC bulk (Fig. \ref{figS4}D, with boundary doing nothing), since they yields the same change of Wilson loop configuration in the spacetime (from Fig. \ref{figS4}A to Fig. \ref{figS4}B).

Therefore, one can imagine the following process which is equivalent to the propagation of chiral Majorana wave packets (Fig. \ref{figS4}C and \ref{figS4}D): when an incident electron on QAH edge A encounters the TSC boundary, one can create two vortices $\sigma_1$ and $\sigma_2$ at the position of the incident electron, then drag the two vortices into the TSC bulk, and trap the incident electron into them at the same time. Similarly, we can create two vortices $\sigma_3$ and $\sigma_4$ at the corner of QAH edge B and drag them into the bulk TSC. Then we braid and fuse the vortices as shown in Fig. \ref{figS4}D, and then drag the fused pair of vortices to QAH edges C and D, respectively. In such a process, the propagation of chiral Majorana fermions on the TSC edge is replaced by braiding of vortices in the bulk, but the outcome remains the same. This shows the two processes are topologically equivalent.

In the end, we briefly clarify the possible conceptual confusions about MZM, Majorana fermion and Ising anyon (vortex). First of all, MZMs or Majorana fermions in any other context (e.g., chiral Majorana fermion on 1D edge) are fermions, and obey fermionic statistics which belongs to \emph{Abelian statistics}. They satisfy the fusion rule $\gamma\times \gamma=1$, namely, the product of two neighbouring Majorana fermion operators gives a topologically trivial bosonic operator. In a topological state of matter, the bosonic operator does not change the topological ground state, thus lives in a 1-dimensional Hilbert space (the ground state), and this means the Majorana fermion operator $\gamma$ is Abelian. Besides, the Majorana fermion operator $\gamma$ satisfy the fermionic statistics that exchanging two fermions yields a phase factor $R_{\gamma\gamma}^1=-1$.

In contrast, the Ising anyons (or vortices) $\sigma$ are non-Abelian anyons. In the bulk of $p+ip$ chiral TSC, $\sigma$ is simply a superconducting vortex where the order parameter $\Delta$ has a $2\pi$ phase winding. They satisfy fusion rules $\sigma\times\sigma=1+\gamma$, which means the product of two nearby $\sigma$ operators can composite into either a bosonic operator $1$ or a fermionic operator $\gamma$. Therefore, two $\sigma$ fields occupy a 2-dimensional Hilbert space, so they obey non-Abelian statistics. In the Ising topological quantum field theory, the braiding of two $\sigma$ fields acquires a phase depending on their fusion channel: when two $\sigma$ are in the fusion channel $1$ and $\gamma$, the braiding phases they acquired are $R_{\sigma\sigma}^1=e^{i\theta}$ and $R_{\sigma\sigma}^\gamma=e^{i\theta+i\pi/2}$, respectively. The two fusion channels thus differ by a $e^{i\pi/2}$ braiding phase.

In the bulk of $p+ip$ TSC, a vortex $\sigma$, namely an Ising anyon, traps a MZM $\gamma$ at the vortex core, which can be seen by solving the Bogoliubov-de Gennes Hamiltonian of the TSC. For this reason, in many discussions the Ising anyon $\sigma$ is not carefully distinguished with the MZM $\gamma$. We emphasize that they are indeed closely related, but are quite different concepts. $\sigma$ is a non-Abelian Ising anyon, while $\gamma$ is Abelian. Their relation can be stated as follows: the \emph{fusion} of two Ising anyons $\sigma_1$ and $\sigma_2$ yields a single fermion degree of freedom, which can be described by the \emph{superposition} of MZM operators $\gamma_1$ and $\gamma_2$.

\end{document}